\documentclass[aps,prl,twocolumn,superscriptaddress,showpacs,floatfix]{revtex4-1}

\usepackage{graphicx}
\usepackage{color}

\usepackage{amsmath,amssymb}

\usepackage{natbib}

\begin{document}

\title{Controlling the dynamics of colloidal particles by critical Casimir forces}

\author{Alessandro Magazz\`u}
\affiliation{Department of Physics, University of Gothenburg, SE-41296 Gothenburg, Sweden}

\author{Agnese Callegari}
\affiliation{Soft Matter Lab, Department of Physics and UNAM -- National Nanotechnology Research Center, Bilkent University, Ankara 06800, Turkey}

\author{Juan Pablo Staforelli}
\affiliation{Center for Optics and Photonics, Universidad de Concepci\'on,  Concepci\'on, Chile}

\author{Andrea Gambassi}
\affiliation{SISSA --- International School for Advanced Studies and INFN, 34136 Trieste, Italy}

\author{Siegfried Dietrich}
\affiliation{Max Planck Institute for Intelligent Systems, Stuttgart, Germany}
\affiliation{IV${}^{\; th}$ Institute for Theoretical Physics, University of Stuttgart, 70569 Stuttgart, Germany}

\author{Giovanni Volpe}
\email{giovanni.volpe@physics.gu.se}
\affiliation{Department of Physics, University of Gothenburg, SE-41296 Gothenburg, Sweden}
\affiliation{Soft Matter Lab, Department of Physics and UNAM -- National Nanotechnology Research Center, Bilkent University, Ankara 06800, Turkey}

\date{\today}

\begin{abstract}
Critical Casimir forces can play an important role for applications in nano-science and nano-technology, owing to their piconewton strength, nanometric action range, fine tunability as a function of temperature, and exquisite dependence on the surface properties of the involved objects. 
Here, we investigate the effects of critical Casimir forces on the free dynamics of a pair of colloidal particles dispersed in the bulk of a near-critical binary liquid solvent, using blinking optical tweezers.
In particular we measure the time evolution of the distance between the two colloids to determine their relative diffusion and drift velocity.
Furthermore, we show how critical Casimir forces change the dynamic properties of this two-colloid system by studying the temperature dependence of the distribution of the so-called first-passage time, i.e., of the time necessary for the particles to reach for the first time a certain separation, starting from an initially assigned one.
These data are in good agreement with theoretical results obtained from Monte Carlo simulations and Langevin dynamics.
\end{abstract}


\maketitle

Critical Casimir forces (CCFs) arise in a binary liquid mixture close to its critical point \cite{Fisher1978,krech94,BTD2000,gambassi2009casimir,gambassi2011critical}. 
Upon approaching the critical point, fluctuations of the composition of the mixture emerge. If these critical fluctuations are confined between neighboring objects (e.g., two colloids, or a colloid and a planar surface), they lead to effective forces between these objects. 
These so-called CCFs were first predicted theoretically in 1978 by M.~E.~Fisher and P.~G.~de Gennes \cite{Fisher1978}
in analogy to quantum-electrodynamical (QED) 
Casimir forces \cite{casimir1948attraction}. 
Only recently they have been measured directly 
\cite{hertlein2008direct,soyka2008critical,paladugu2016nonadditivity} 
and proved to be relevant for soft matter \cite{bonn2009direct,commentBonn,piazza2011depletion}.
These CCFs have been enjoying significant interest both from basic research and because they are promising candidates for applications in nano-science and nano-technology, 
in order to manipulate objects (e.g., by controllable periodic deformations of chains), to assemble devices (e.g., via the self-assembly of colloidal molecules \cite{marino2016assembling,nguyen2017tuning}), and to drive machines (e.g., by powering rotators \cite{schmidt2017microscopic}) at the nano- and micro-meter scale.
In fact, their piconewton strength and nanometric ranges of action match the requirements of nano-technology.
Furthermore, these forces show an exquisite dependence on the temperature of the environment and on the chemical surface properties of the objects involved \cite{gambassi2011critical,gambassi2009casimir,soyka2008critical,nellen2009tunability,pattexp2011}. For example, if density fluctuations are confined between particles with the same surface property (e.g., hydrophilic), attractive CCFs take hold, while they are repulsive between particles with opposite surface properties (e.g., hydrophilic vs. hydrophobic particles).
With the exception of Ref.~\cite{mart2017}, until now, the experimental studies have focused on the time-independent properties of CCFs and thus keeping their dynamics hidden. 
Here, we use blinking optical tweezers to reveal how CCFs affect the free dynamics of a pair of colloidal particles immersed in a binary solution.

\begin{figure*}[t] 
\includegraphics [width=\textwidth]{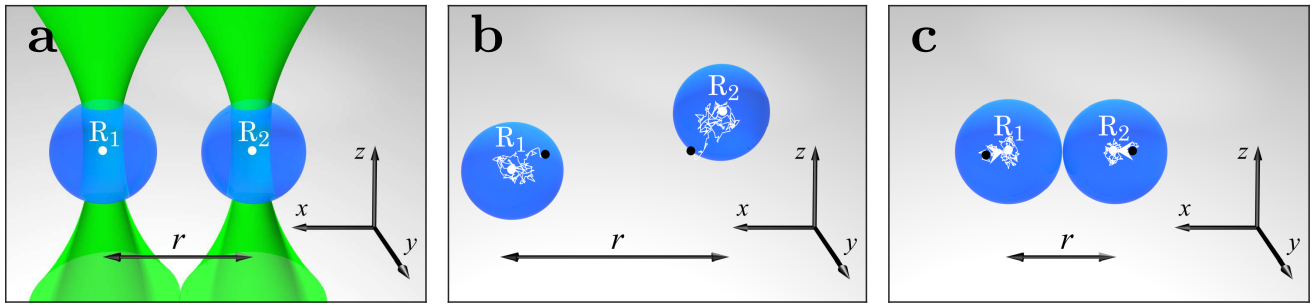}
\caption{Schematic presentation of the design of the experiment.
(a) Two equal spherical silica colloids (blue spheres, diameter $d=2.06\pm0.05\,{\rm \mu m}$) are optically trapped by two traps (green laser beams focused on the points $\mathbf{R}_{0,1}$ and  $\mathbf{R}_{0,2}$) in the bulk of a binary liquid mixture of water and 2,6-lutidine.  
The centers of the two colloids shown in the figure are located at positions $\mathbf{R}_{1,2}$ (rather close to $\mathbf{R}_{0,1}$ and $\mathbf{R}_{0,2}$), and the lateral distance between them, i.e., projected onto the $xy$-plane which the laser beams are orthogonal to, is indicated by the black arrowed line $r$.
(b) If the temperature of the mixture $T$ is sufficiently far away from the critical temperature $T_{\rm c}$ (i.e., $T_{\rm c}-T=\Delta T \gtrsim 500\,{\rm mK}$), upon switching off the optical tweezers by blocking the laser beam, the two colloids start to diffuse \emph{freely} in the solvent. 
The positions of the colloids change from their initial values $\approx \mathbf{R}_{0,1}$ and $\approx \mathbf{R}_{0,2}$, indicated by the black dots, to the final ones $\mathbf{R}_1$ and $\mathbf{R}_2$, respectively, indicated by the white dots, following the irregular white trajectory.
(c) As $T$ approaches $T_{\rm c}$, attractive CCFs (white arrows) arise and affect the dynamics of the particles such that there is a large probability for them to approach each other.
}
\label{fig1}
\end{figure*}

\begin{figure}[h!]
\includegraphics [width=0.5\textwidth]{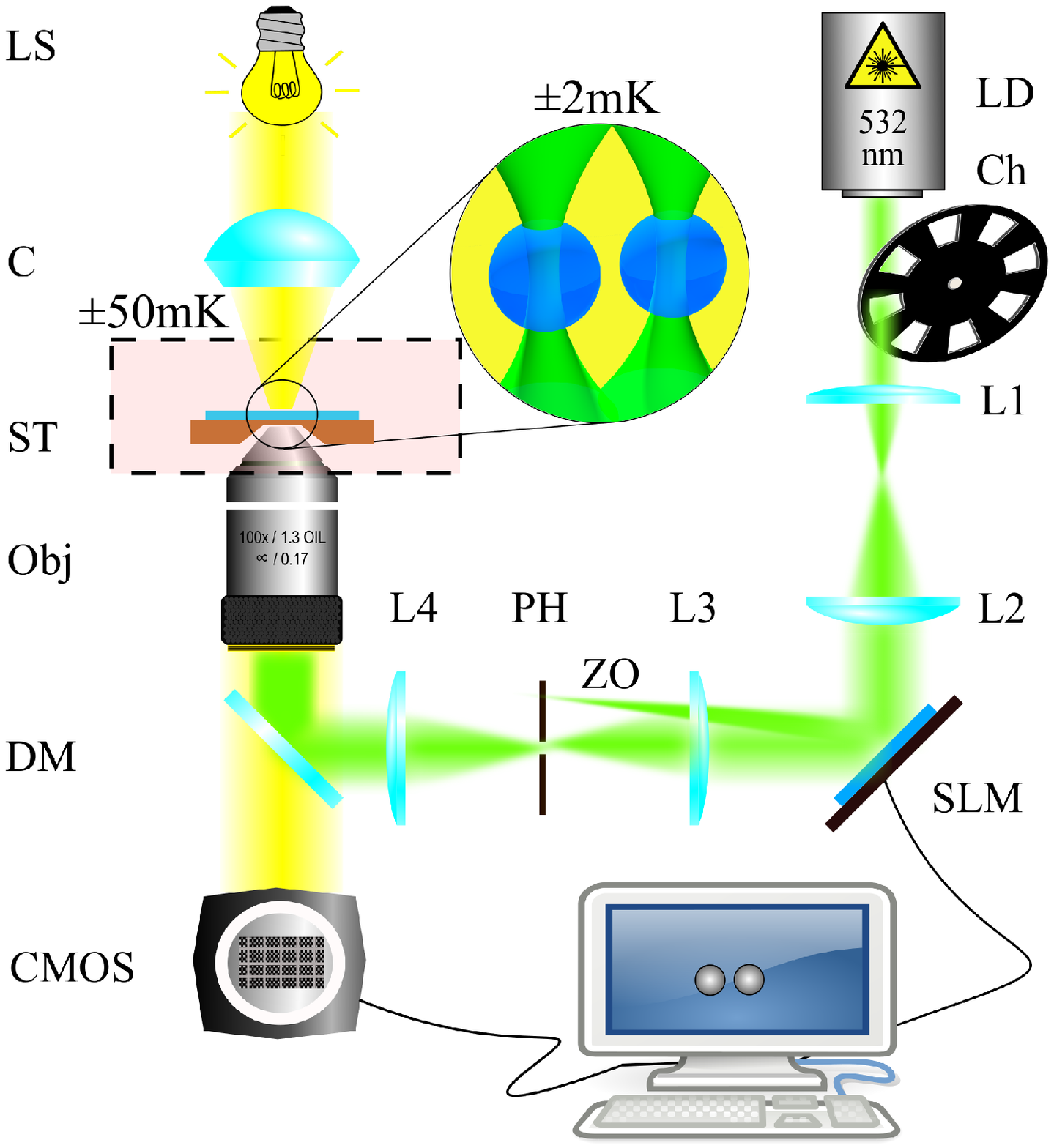}
\caption{Scheme of the experimental setup.
The experimental setup consists of three main parts: holographic optical tweezers, a digital video microscope, and a temperature control unit. 
The holographic optical tweezers generate two optical traps by imposing a phase-only hologram on an incoming laser beam by using a spatial light modulator (SLM) and by focusing the resulting beam through an oil-immersion objective (Obj) (with magnification $100\times$ and numerical aperture NA = 1.30).
The laser beam with a  wavelength of $532\,{\rm nm}$ is generated by a laser diode (LD) and the traps are periodically switched on and off by a chopper (Ch). A telescope consisting of two lenses (L1 and L2) is used to overfill the active area of the SLM. Two other lenses (L3 and L4) and a pinhole (PH) are used for blocking the zeroth order (ZO) of the reflected beam. 
The digital video microscope is used to track the positions of the colloidal particles by illuminating the sample with a white light source (LS), focused by a condenser (C). A dichroic mirror (DM) is used to combine the optical paths of the laser and of the white light. The entire trapping process is imaged by a CMOS camera. 
The temperature of the whole sample stage (ST) is stabilised with a precision of $50\,{\rm mK}$. It is located inside an enclosed chamber indicated by the shaded, pink area with a dashed contour. The temperature of the focal region (inset) is measured and stabilised to within $\pm2\,{\rm mK}$ through the objective with a feedback controller.}
\label{fig2}
\end{figure}

\emph{Experimental setup and data analysis --} In our experiment we use silica microspheres with diameter $d=2.06\pm0.05\, {\rm \mu m}$ (Microparticles GmbH), dispersed in a critical mixture of water and 2,6-lutidine at the critical lutidine mass fraction $c^{\rm c}_{\rm L}=0.286$, corresponding to a lower critical point at the temperature $T_{\rm c} \simeq 34^\circ {\rm C}$ \cite{grattoni1993lower,gambassi2009critical}.
In the bulk of the critical mixture, we generate two holographic optical traps \cite{grier2003revolution,jones2015optical} at positions $\mathbf{R}_{0,1}$ and $\mathbf{R}_{0,2}$ in order to fix the positions $\mathbf{R}_1$ and $\mathbf{R}_2$ of the centers of two spherical colloids at their initial values, approximately equal to $\mathbf{R}_{0,1}$  and $\mathbf{R}_{0,2}$, respectively, with a specified center-to-center distance $r_0 = 2.40\,{\rm \mu m}$ (Fig.~\ref{fig1}a).
We have chosen this value for $r_0$ such that the resulting surface-to-surface distance between the two colloids is of the order of $300\,{\rm nm}$. Accordingly, the latter is significantly larger than the range of the electrostatic repulsion (which holds up to a surface-to-surface distance of about $100\,{\rm nm}$, because the Debye screening length is $\ell_{\rm D}\simeq 13\,{\rm nm}$ \cite{paladugu2016nonadditivity,gambassi2009critical}) and comparable with the largest range of the CCFs achieved in the present experiment (which here are always negligible beyond $\approx 300\,{\rm nm}$).
Given the sensitivity of CCFs to temperature, the sample temperature is measured and stabilized with a feedback controller to within $\pm2\,{\rm mK}$ \cite{paladugu2016nonadditivity}. The scheme of the experimental setup is presented in Fig.~\ref{fig2}.

We periodically chop the laser beam at the blinking frequency $f_{\rm b} = 1.3\,\,{\rm Hz}$ so that the optical traps are periodically switched on and off (blinking optical tweezers \cite{grier1997optical,pesce2010blinking,pesce2014long}).
We have chosen this value of $f_{\rm b}$ because it is sufficiently low in order to be able to observe the effects of CCFs on the particle dynamics in the $xy$-plane and high enough to permit us to neglect the effects of gravity on the vertical position $z$ of the particle.
We record the ensuing motion of the colloids at $300\,{\rm fps}$ during the time windows in which the beam is blocked and hence the optical potential is not present.
When the traps are turned on again, the two colloids are brought back to their initial positions by the restoring forces of the optical potentials. 
If the temperature $T$ of the mixture is sufficiently far from $T_{\rm c}$ (i.e., $\Delta T=T_{\rm c}-T \gtrsim 500\,{\rm mK}$), the two particles {\em freely} diffuse in the solution as long as the optical traps are off (Fig.~\ref{fig1}b). 
When $\Delta T \to 0$, critical order parameter fluctuations take hold associated with an increasing correlation length $\xi$. As $\xi$ becomes comparable to the inter-particle distance $r$, the two hydrophilic particles experience attractive CCFs, which affect their dynamics and reduce their inter-particle distance (Fig.~\ref{fig1}c).
The entire blinking process is repeated about 400 times for each fixed value of $\Delta T \to 0$ in order to acquire sufficient statistics for the dynamics of the colloids.

We have analyzed the acquired videos using digital video microscopy \cite{crocker1996methods,jones2015optical} in order to determine the trajectories $\mathbf{r}_{1}(t)$ and $\mathbf{r}_{2}(t)$ of the centers of the two particles projected onto the $xy$-plane, where $\mathbf{r}_{l}(t) = (x_{l}(t), y_{l}(t))$ with  $l=1, 2$ labelling the particles.
We correct the relative position of the particles $\mathbf{r}(t)= \mathbf{r}_{\rm 2}(t)-\mathbf{r}_{\rm 1}(t) $ and their relative distance $r(t)=|\mathbf{r}(t)|$ to account for artefacts which appear in digital video microscopy due to the proximity between the two particles \cite{baumgartl2005limits,paladugu2016nonadditivity}.

\begin{figure*}[t] 
\includegraphics[width=0.9\textwidth]{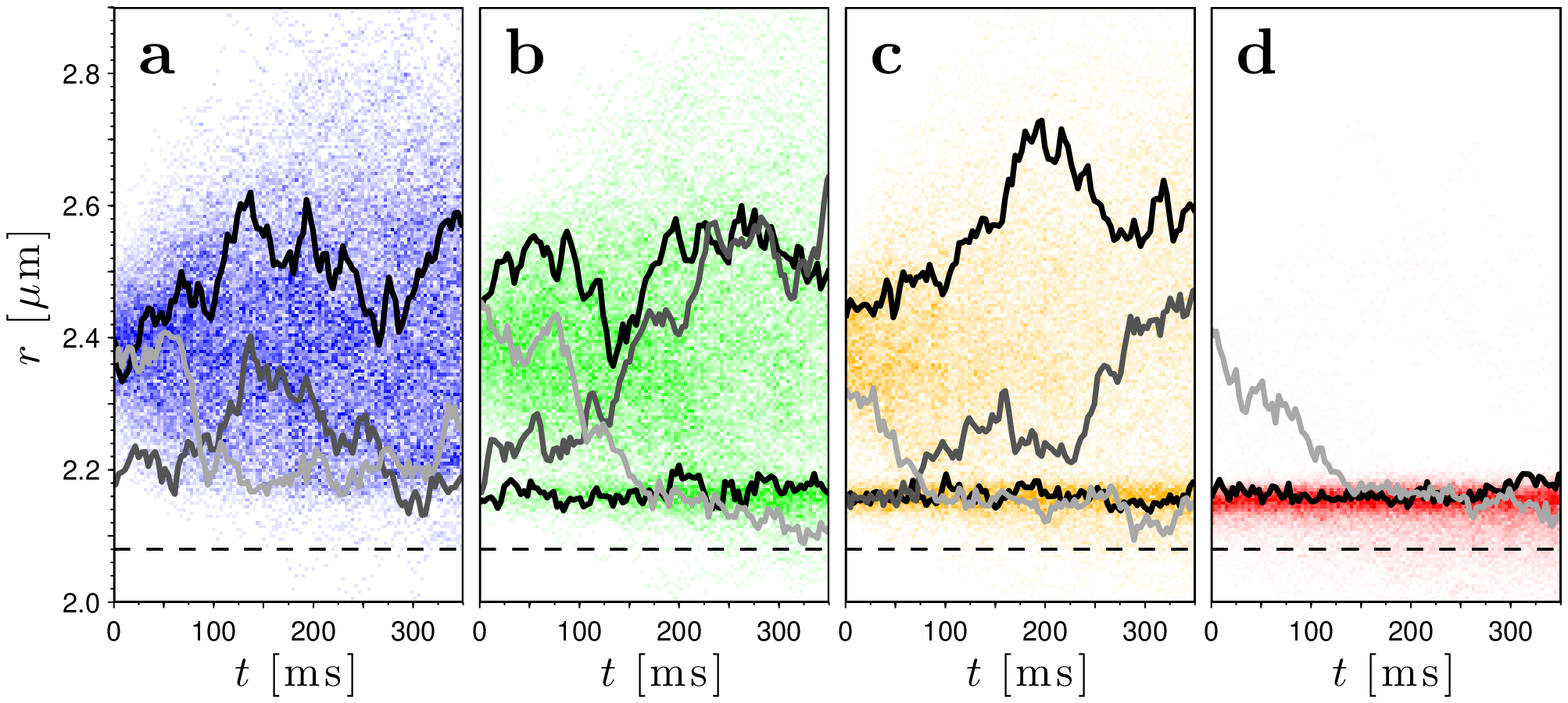}
\caption{Time evolution of the projected inter-particle distances $r(t)$ after the optical tweezers are switched off at $t=0\,{\rm ms}$ for decreasing values of $\Delta T$: (a) $\Delta T=456\pm2\,{\rm mK}$, (b) $200\pm2\,{\rm mK}$, (c) $163\pm2\,{\rm mK}$, and (d) $108\pm2\,{\rm mK}$. The solid lines indicate representative individual trajectories while the intensities of the background colors represent the evolution of the particle position probability distribution obtained from 400 different trajectories.
From (a) to (d), the free diffusion of the colloids is increasingly affected by the emergence of attractive CCFs upon approaching criticality.
The dashed horizontal line indicates the projected inter-particle distance $r$ corresponding to the diameter $d$ of the colloids. Sometimes $r(t)$ is smaller than $d$ because a displacement of the colloids along the vertical $z$-axis causes their projections onto the $xy$-plane to overlap.
}
\label{fig3}
\end{figure*}

\emph{Particle trajectories and time-evolution of the inter-particle distance probability density --} In Fig.~\ref{fig3} several inter-particle projected distances $r(t)$ are reported  for decreasing values of $\Delta T$. For each value of $\Delta T$, we plot the evolution of the inter-particle distance probability density as a function of time obtained from 400 particle trajectories by binning (colored background). We highlight a few selected trajectories to illustrate typical behaviors of the particles (solid lines).
For $\Delta T=456\pm2\,{\rm mK}$ (Fig.~\ref{fig3}a), the particles are diffusing freely and the CCFs do not affect their behavior. This can be inferred from the fact that the inter-particle distance probability density is rather broad.
When $\Delta T$ is reduced to $\Delta T = 200\pm2\,{\rm mK}$ (Fig.~\ref{fig3}b) and $163\pm2\,{\rm mK}$ (Fig.~\ref{fig3}c), the CCFs arise and affect the dynamics of the colloids. Occasionally they cause adhesion as can be inferred from the emergence of a peak in the inter-particle distance probability density at $r\approx 2.16\,{\rm \mu m}$.
If $\Delta T$ is reduced further to $\Delta T=108\pm2\,{\rm mK}$ (Fig.~\ref{fig3}d), strong attractive CCFs hinder the free diffusion of the particles, which often adhere to each other so that the values of $r$ lie within a small region in which there is a balance between the repulsive electrostatic forces and the attractive CCFs.

\begin{figure*}[t]
\includegraphics[width=\textwidth]{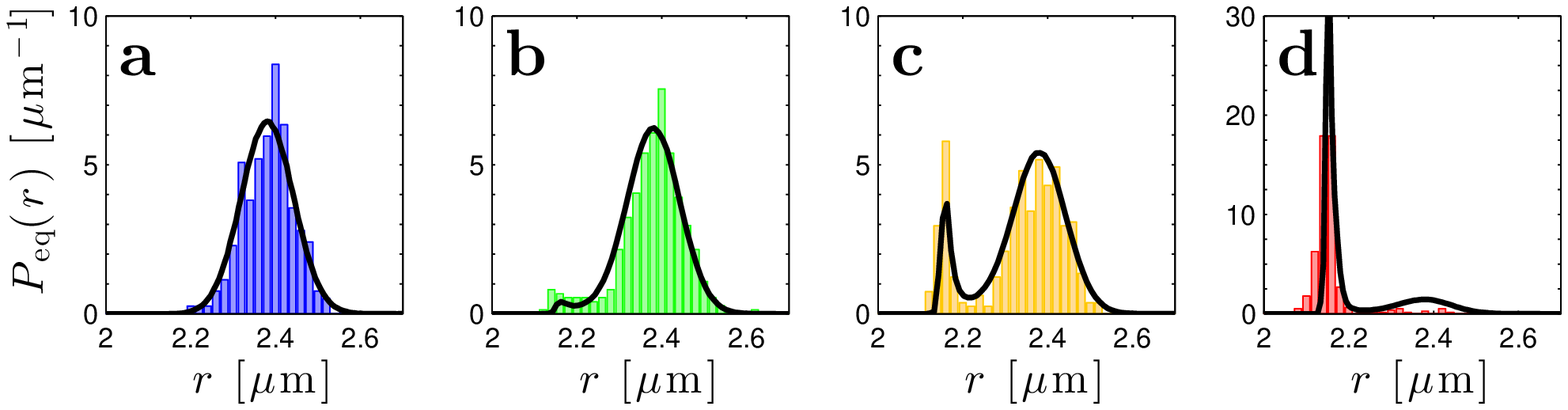}
\caption{Equilibrium distribution $P_{\rm eq}(r)$ of the inter-particle distance $r(0)$ (i.e., when the optical tweezers are switched off) for two optically trapped colloids at temperatures (a) $\Delta T=456\pm2\,{\rm mK}$, (b) $200\pm2\,{\rm mK}$, (c) $163\pm2\,{\rm mK}$, and (d) $108\pm2\,{\rm mK}$.
Each histogram is obtained from 400 different experimental values. 
The solid black lines are the theoretical  distribution of $r(0)$, obtained via Monte Carlo integration ($10^6$ samples) of two optically trapped particles subjected to the theoretical total potential $V(\mathbf{R}_1,\mathbf{R}_2)$ (Eq.~\eqref{eq:2}).
}
\label{fig4}
\end{figure*}

\emph{Equilibrium distributions and parameter fitting --} At a specific temperature $\Delta T$, the initial values $\mathbf{r}_{1,2}(0)$ of the recorded trajectories are sampled from the equilibrium distribution of the two trapped particles exposed to the optical potentials $V_{{\rm ot},1}(\mathbf{R}_1)+V_{{\rm ot},2}(\mathbf{R}_2)$, to the electrostatic repulsion $V_{\rm es}(\rho)$, and possibly to CCFs $V_{\rm C}(\rho)$, i.e., to the total potential
\begin{equation}\label{eq:2}
V(\mathbf{R}_1,\mathbf{R}_2)=V_{{\rm ot},1}(\mathbf{R}_1)+V_{{\rm ot},2}(\mathbf{R}_2)+V_{\rm es}(\rho)+V_{\rm C}(\rho),
\end{equation}
where $\rho = |\mathbf{R}_2-\mathbf{R}_1| - d$ is the actual surface-to-surface distance between the two colloids, and $\mathbf{R}_l = (x_l,y_l,z_l)$ with $l=1,2$ are their initial positions. Note that the projected distance $r$ introduced above is generically smaller than the actual center-to-center distance $|\mathbf{R}_2-\mathbf{R}_1|$, due to possible displacements of the colloids along the vertical $z$ direction.
During the time windows when the optical traps are switched on, the positions of the particles evolve under the action of the total potential $V(\mathbf{R}_1,\mathbf{R}_2)$, and, after a sufficiently long time, they reach the equilibrium distribution $P_{\rm eq}(\mathbf{R}_1,\mathbf{R}_2) \propto \exp[-V(\mathbf{R}_1,\mathbf{R}_2)/(k_{\rm B}T)]$.
Accordingly, when the optical traps are switched off, the distribution of the initial positions of the particles renders $P_{\rm eq}(r)$, which follows from $P_{\rm eq}(\mathbf{R}_1,\mathbf{R}_2)$ 
by integrating over all the possible configurations with projected center-to-center distance  equal to $r$, i.e., $P_{\rm eq}(r) = \int {\rm d}^3 \mathbf{R}_{1}\,{\rm d}^3 \mathbf{R}_{2} \,  P_{\rm eq}(\mathbf{R}_{1},\mathbf{R}_{2})\,\delta\left(r-\left|\mathbf{r}_{1} - \mathbf{r}_{2}\right|\right)$, where we remind that $\mathbf{r}_l$ is the projection of $\mathbf{R}_l$ onto the $xy$-plane.
These distributions are reported in Fig.~\ref{fig4} for the same values of $\Delta T$ as  considered in Fig.~\ref{fig3}. The histograms represent the experimentally measured data and the solid lines are the corresponding theoretical results obtained from 
the Monte Carlo integration of $P_{\rm eq}(\mathbf{R}_{1},\mathbf{R}_{2})$, introduced  above.

For $\Delta T = 456\pm2\, {\rm mK}$ (Fig.~\ref{fig4}a), $P_{\rm eq}(r)$ can be very well approximated by a Gaussian distribution centered at the value $r\simeq2.40\,\,{\rm \mu m}$,  corresponding to the distance $r_0$ between the minima of the two optical potentials. This result is expected for two optically trapped particles which do not interact with each other \cite{jones2015optical}.
Reducing $\Delta T$ (Figs.~\ref{fig4}b-d), a peak arises at $r\simeq2.16\,{\rm \mu m}$ on the left flank of the Gaussian distribution, becoming more dominant at the expense of the Gaussian distribution. This is due to the gradual emergence of attractive CCFs between the particles, which causes them to adhere to each other  also in the presence of the optical potentials. The peak position indicates the region where the repulsive electrostatic forces and the attractive CCFs are balanced.

By using Monte Carlo integration we also calculated the distribution $P_{\rm eq}(r)$ on the basis of the theoretical potential $V(\mathbf{R}_1,\mathbf{R}_2)$. Adjusting the parameters of the theoretical potential, it is possible to match the experimental distribution $P_{\rm eq}(r)$ at $t=0$. By doing so, we have obtained, for each value of $\Delta T$, the correlation length $\xi$ of the order parameter fluctuations \cite{paladugu2016nonadditivity} and the Debye screening length $\ell_{\rm D}$ of the electrostatic interaction as fit parameters. 
Specifically, we adopt for $V_{\rm C}$ the theoretical prediction
\begin{equation}\label{eq:VC}
V_{\rm C}(\rho) = k_{\rm B} T_c \,\frac{d}{4\rho} \Theta(\rho/\xi), 
\end{equation}
based on the Derjaguin approximation (see, e.g., Ref.~\cite{gambassi2009critical}).
$\Theta$ is a scaling function, which can be inferred from the numerical estimates available in the literature \cite{vas2007MC,vas2009MC,hasenbusch2012thermodynamic,paladugu2016nonadditivity}.
For $V_{\rm es}$, we consider the simple expression \cite{paladugu2016nonadditivity,gambassi2009critical}
\begin{equation}\label{eq:Ves}
V_{\rm es}(\rho) = k_{\rm B} T_{\rm c} {\rm e}^{- (\rho-\rho_{\rm es})/\ell_{\rm D}},
\end{equation}
where $\rho_{\rm es}$ depends on the surface charges of the colloids and is used here as a fit parameter, with $\rho_{\rm es} \simeq 95\,{\rm nm}$ and  $\ell_{\rm D}\simeq 13\,{\rm nm}$, which are in line with the values found in previous investigations \cite{paladugu2016nonadditivity,gambassi2009critical}. 
The optical potentials $V_{{\rm ot},1}$ and $V_{{\rm ot},2}$ are assumed to be harmonic, i.e., 
\begin{equation}
V_{{\rm ot},l}(\mathbf{R}_l)=\frac{1}{2}k_l(\mathbf{R}_l - \mathbf{R}_{0,l})^2,
\label{eq:Vopt}
\end{equation}
where $\mathbf{R}_{0,l}$ is the center of trap $l=1,2$ and the  stiffness $k_{1,2}$ is determined experimentally for each trap.
Then, these parameters are fixed to their best-fit values and the resulting total potential is used to simulate the ensuing evolution by Brownian dynamics, taking into account the distribution of the initial conditions. 
The core of a Brownian dynamics simulation is given by the Langevin equation, which is a stochastic differential equation describing the time evolution of a particle performing Brownian motion. It  is integrated in time in order to create trajectories of the particle \cite{volpe2013simulation}.
The values of $T$ and $T_{\rm c}$ are obtained by fitting the experimental values of $\xi$ for various $\Delta T$ to the theoretical prediction \citep{paladugu2016nonadditivity}
\begin{equation}\label{xi}
\xi(T)=\xi_0\left ( 1-\frac{T}{T_{\rm c}}\right )^{\nu},
\end{equation}
where the non-universal length $\xi_0=0.20\pm0.02 \,{\rm nm}$ has been determined by light-scattering experiments for the water-2,6-lutidine mixture \citep{gulari1972light}, and $\nu=0.63$ is a universal bulk critical exponent holding for classical binary liquid mixtures \citep{gambassi2009critical}.
We emphasise that, while the optical tweezers are switched off, i.e., with $V_{{\rm ot},l}=0$, the particles undergo Brownian motion and diffuse under the influence of the CCFs (due to $V_{\rm C}$) and of the electrostatic interaction (due to $V_{\rm es}$).

\emph{Diffusion and drift velocity --} The relative position and the distance between the two particles can be used to determine the values of their diffusion constant $D(r)$ and of their drift velocity $v(r)$ as functions of $r$ \cite{brettschneider2011force}. 
Given the experimental trajectories of the particles as a sequence of values $r_i$, numbered by $i$ and acquired at times $i\times t_{\rm s}$, where $t_{\rm s}$ is the time between sampling, one has
\begin{equation}\label{dif}
D(r) = \frac{1}{2} \left\langle  \left. \frac{(r_{i+n}-r_i)^2}{n t_{\rm s}} \right|  r_i \in \left\lbrack r-\delta r, r+\delta r  \right\rbrack \right\rangle
\end{equation}
and
\begin{equation}
\label{drif}
v(r) = \left\langle \left. \frac{r_{i+n}-r_i}{n t_{\rm s}} \right|   r_i \in \left\lbrack r-\delta r, r+\delta r  \right\rbrack \right\rangle,
\end{equation}
where $\langle  \ldots | \ldots \rangle$ denotes the average over the various trajectories under the specified condition; $\delta r=10\,{\rm nm}$ is the spatial resolution for $D(r)$ and $v(r)$, with $n=3$ for determining the diffusion coefficient and $n=10$ for the drift. 
These values of $n$ have been chosen separately as the smallest integers which render statistically meaningful, almost $n$-independent  values for $D(r)$ and $v(r)$.
In particular, in oder to estimate the actual diffusion coefficient via the estimator $D$, it would be desirable to consider small values of $n$ in such a way that the potential effects of a deterministic drift are negligible (see also the discussion below). On the contrary, for the calculation of $v$, it would be preferable to consider larger values of $n$, in order for the effects of diffusion to average out. However, exceedingly small values of $n$ yield numerical data for $D$ with large statistical fluctuations, while exceedingly large values of $n$ would not allow for a proper identification of the spatial dependence of $v(r)$. The choices indicated above emerge from a compromise between these competing requests.

In Eq.~\eqref{dif} [and in, c.f., Eqs.~\eqref{eq:def-Dpp} and \eqref{eq:def-Dpe}]  the actual diffusion constant is determined from the experimental data via the estimator  $D = \langle (\Delta r)^2 \rangle/(2 \Delta t)$, in terms of the (conditional) average of the displacement $\Delta r$ observed within a suitable time interval $\Delta t$ ($\, = n\, t_{\rm s}$ in 
Eq.~\eqref{dif}). However, an alternative estimator for the same quantity is $\hat D = [\langle (\Delta r)^2 \rangle - \langle \Delta r\rangle^2]/(2 \Delta t)$, which is closer to the common definition of the diffusion constant and which highlights the sole effect of the Brownian noise, as it subtracts a possible mean drift $\langle \Delta r\rangle$ of the particles due to the action of external forces in the presence of overdamped dynamics. 
These two estimators are actually related by
$\hat D = D - \Delta t \, v^2/2 + {\cal O}((\Delta t)^2)$,
where $v = \langle \Delta r\rangle/\Delta t$ is the average relative velocity of the particles, and therefore they render the same value for sufficiently small $\Delta t$ or whenever $v$ vanishes  due to the absence of external forces. 
In the present experiment we considered the estimator $D$ instead of $\hat D$ for three reasons: (a) Since $\langle \Delta r \rangle$, i.e., $v\, \Delta t$ is affected by statistical errors, subtracting it from  $\langle (\Delta r)^2 \rangle$ increases the resulting statistical error of $\hat D$ compared to that of $D$; (b) At distances $r$ larger than ca.~$2.2 \mu{\rm m}$, we expect all deterministic forces involved to vanish in the absence of the tweezers and therefore $v=0$, such that $D = \hat D$; (c)  At smaller distances, but for sufficiently small values of $\Delta t$, i.e., for $\Delta t \ll D/v^2$, the two estimators $\hat D$ and $D$ are anyhow approximately equal.
Both estimators $D(r)$ and $\hat D(r)$ render the actual diffusion coefficient as $\Delta t \to 0$ but they are generically affected by (different) finite-time corrections which depend on $r$, are linear in $\Delta t$ for small $\Delta t$, and are related as discussed above \cite{fn1}.
\begin{figure}[t]
\includegraphics[width=0.45\textwidth]{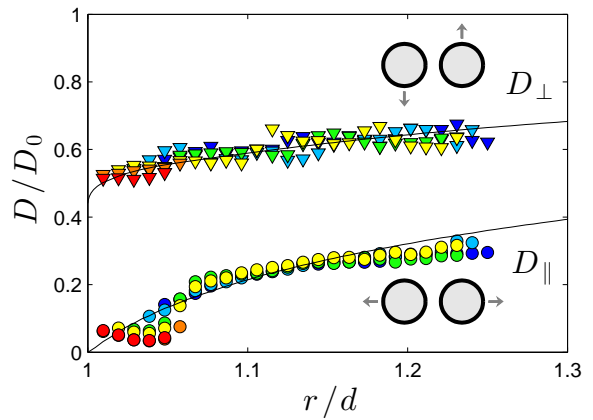}
\caption{Experimental values of the normalized parallel $D_\| / D_0$ (circles) and perpendicular $D_\perp / D_0$ (triangles) diffusion constants as functions of the ratio between the inter-particle center-to-center distance $r$ and the particle diameter $d$.
The parallel and perpendicular directions refer to the line connecting the centers of the two colloids, and $D_0$ is the bulk diffusion constant determined by Eq.~\eqref{eq:D0} and from available experimental data.
The various colors refer to data taken at $\Delta T=456\pm2\,{\rm mK}$ (blue),  $273\pm2\,{\rm mK}$ (light blue), $200\pm2\,{\rm mK}$ (green), $163\pm2\,{\rm mK}$ (yellow), $127\pm2\,{\rm mK}$ (orange), and $108\pm2\,{\rm mK}$ (red).
The solid lines represent the theoretical predictions accounting for the effect of the hydrodynamic interaction between the colloids \cite{batchelor1976brownian}. 
There is good agreement between the theoretical predictions and the experimental results, which neither exhibit an appreciable dependence on temperature. 
The nature of the deviations observed in $D_\|$ at short distances is discussed in the main text.
}
\label{fig5}
\end{figure}

Due to the hydrodynamic interaction between the two colloids, the actual diffusion constant of these particles differs from the free one they would have in the bulk \cite{batchelor1976brownian}. As a result, the diffusion along the direction connecting the centers of the particles occurs with a diffusion constant which differs from that in the direction perpendicular to it (see Eq.~(5.5) in Ref.~\cite{batchelor1976brownian}).
In order to compare our experimental results with these theoretical predictions, we decompose the $i$-th displacement $\Delta\mathbf{r}^{(n)}_{i} = \mathbf{r}_{i+n} - \mathbf{r}_{i}$ into its parallel and perpendicular components:
\begin{equation}
\Delta r^{(n)}_{i \parallel} = \Delta\mathbf{r}^{(n)}_{i} \cdot \hat{\mathbf{r}}_{i}
\end{equation}
and 
\begin{equation}
\Delta r^{(n)}_{i \perp} = \Delta\mathbf{r}^{(n)}_{i} \cdot (\hat{\mathbf{z}} \times \hat{\mathbf{r}}_{i}),
\end{equation}
where $\hat{\mathbf{r}}_{i}=\mathbf{r}_{i}/r_{i}$  and $\hat{\mathbf{z}}$ is the unit vector along the $z$-direction, which is perpendicular to the $xy$-plane of observation where the position vectors $\mathbf{r}_{i}$ lie.
With these definitions, we can obtain the parallel and perpendicular diffusion coefficients  defined with respect to the direction connecting the centers of the two particles:
\begin{equation}
\label{eq:def-Dpp}
D_{\parallel} (r) = \frac{1}{2} \left\langle  \frac{|\Delta {r}^{(n=3)}_{i \parallel}|^2}{3\times t_{\rm s}} \left| \frac{}{} \right. r_{i} \in \left\lbrack r-\delta r, r+\delta r  \right\rbrack \right\rangle,  
\end{equation}
and
\begin{equation}
\label{eq:def-Dpe}
D_{\perp} (r) = \frac{1}{2} \left\langle  \frac{|\Delta {r}^{(n=3)}_{i \perp}|^2}{3\times t_{\rm s}} \left| \frac{}{} \right. r_{i} \in \left\lbrack r-\delta r, r+\delta r  \right\rbrack \right\rangle,
\end{equation}
respectively, and the parallel and perpendicular drift velocities:
\begin{equation}
\label{eq:def-vpp}
v_{\parallel}(r) = \left\langle  \frac{\Delta {r}^{(n=10)}_{i \parallel}}{10\times t_{\rm s}} \left| \frac{}{} \right.  r_{i} \in \left\lbrack r-\delta r, r+\delta r  \right\rbrack \right\rangle,  
\end{equation}
and
\begin{equation}
\label{eq:def-vpe}
v_{\perp}(r) = \left\langle  \frac{\Delta {r}^{(n=10)}_{i \perp}}{10\times t_{\rm s}} \left| \frac{}{} \right.  r_{i} \in \left\lbrack r-\delta r, r+\delta r  \right\rbrack \right\rangle,
\end{equation}
respectively.
The remark about the use of the estimator $D$ instead of $\hat D$, stated after Eq.~\eqref{dif}, here applies to $D_\|$, while $D_\bot$ is not affected by the choice of the estimator because no forces are expected to act along the direction perpendicular to the segment joining the centers of the colloids.
In Fig.~\ref{fig5} we report the experimental data (symbols) for $D_\perp$ (upper set) and $D_\|$ (lower set) as functions of the ratio $r/d$, together with the theoretical prediction obtained in Ref.~\cite{batchelor1976brownian} (solid line) for no-slip boundary conditions.  These quantities are normalized by the bulk inter-particle diffusion constant given 
by (see, e.g., Eq.~(5.6) in Ref.~\cite{batchelor1976brownian})
\begin{equation}
\label{eq:D0}
D_0 = \frac{k_{\rm B} T}{3 \pi \eta \, d/2}  \simeq 0.22 \, (\mu{\rm m})^2/{\rm s},
\end{equation} 
in terms of the viscosity $\eta \simeq 2 \times 10^{-3}\,{\rm Ns/m^2}$ of the mixture close to $T_{\rm c}$ \cite{clunie1999interdiffusion}. 
(The expected singularity of $D_0$ and $\eta$ upon approaching criticality is so mild that it can be neglected for all practical purposes \cite{eta72,eta89}.)
Both for $D_{\perp}$ and $D_{\parallel}$ Fig.~\ref{fig5} shows satisfactory agreement, with a systematic discrepancy emerging only in $D_\|$ for $r/d \lesssim 1.05$.
This discrepancy is due to the limited experimental acquisition rate, which does not allow us to resolve times shorter than $3\,{\rm ms}$. In fact, the same discrepancy is encountered 
in simulations when we consider trajectories sampled with the same time step  $t_{\rm s}$ as the one used in the experiment. If we reduce significantly the time step $t_{\rm s}$ in the simulations, the discrepancy with the theoretical line is much less pronounced  and eventually disappears, as it is expected from the fact that the latter should be recovered as $t_{\rm s} \to 0$.
In addition, we note that both in experiment and simulation, values of $r/d$ which are smaller than 1.05  are obtained only when the temperature $T$ is sufficiently close to $T_c$ and, thus, sizable CCFs are present, especially at such short distances. Their presence implies $v\neq 0$ and therefore we expect finite-time corrections to appear in $D_\|\simeq \hat D_\| + v^2_\| \Delta t/2$ in addition to those which characterize  $\hat D_\|$.
The experimental data for $D_{\perp,\|}$ reported in Fig.~\ref{fig5} do not show any significant dependence on $\Delta T$ and $\xi$, apart from the finite-time effects mentioned above. 
A genuine temperature dependence of $D_{\perp,\|}$ could be expected for Brownian particles diffusing near $T_{\rm c}$ in an external potential, provided by strongly temperature-dependent critical fluctuations which alter the dynamics (see, e.g., Refs.~\cite{Fuji17,Fuji16} for a single trapped colloid.)  Accordingly, the absence of this dependence suggests that the effective interaction $V_{\rm C}$ is valid, for all practical purposes, as if the colloidal particles were at rest in their instantaneous position. 
This also implies that for the present experimental conditions the effects of \emph{retardation}, observed numerically in 
Ref.~\cite{furukawa2013nonequilibrium} during the aggregation of two identical colloids due to CCFs, are negligible.

\begin{figure*}[h!]
\includegraphics[width=0.95\textwidth]{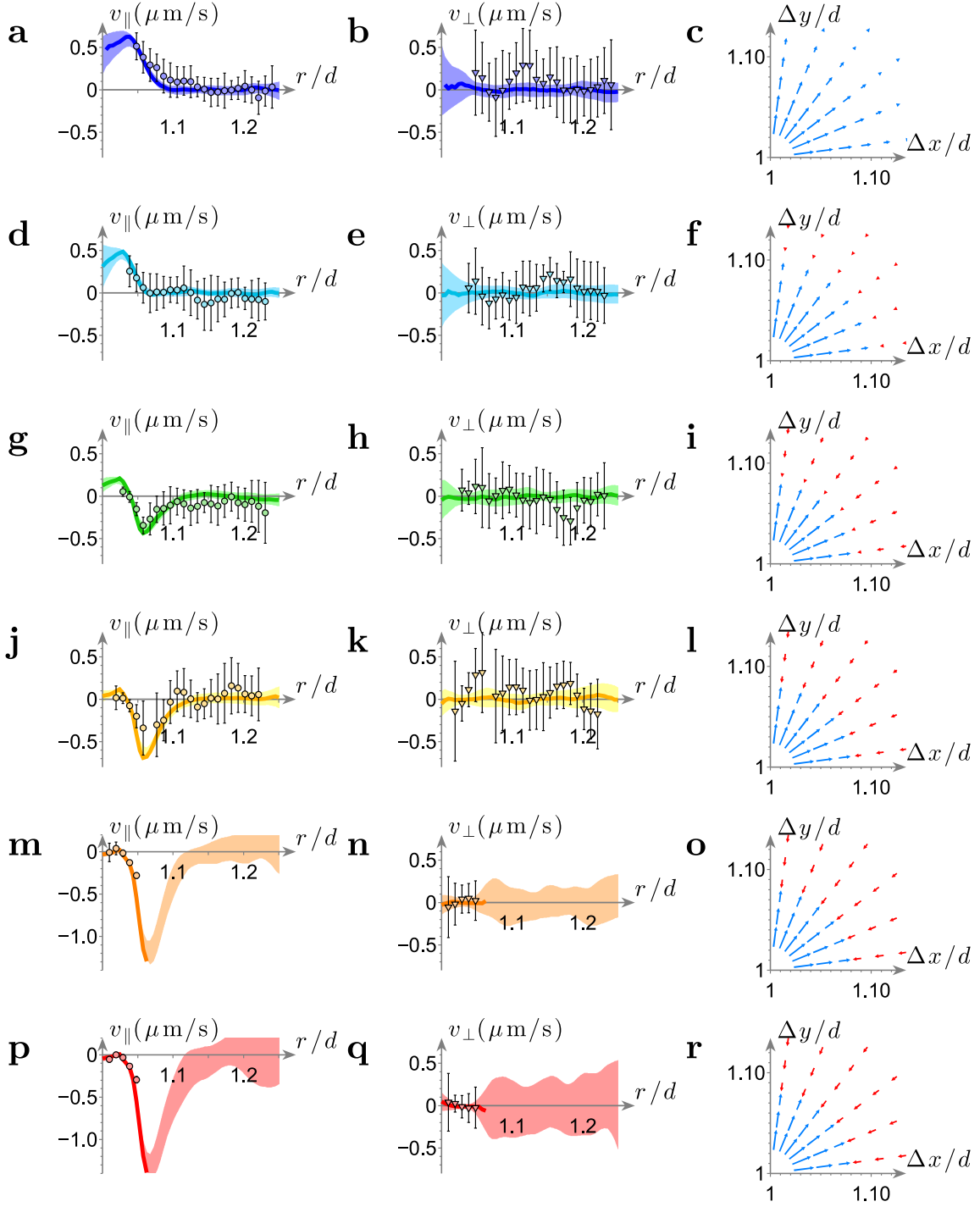}
\caption{The drift velocities $v_\parallel$ (first column) and $v_\perp$ (second column), parallel and perpendicular, respectively, to the direction connecting the centers of the two colloids and the total force field (in arbitrary units) used as the basis of the simulations (third column) are represented for (a-c) $\Delta T=456\pm2\,{\rm mK}$, (d-f) $273\pm2\,{\rm mK}$, (g-i) $200\pm2\,{\rm mK}$, (j-l) $163\pm2\,{\rm mK}$, (m-o) $127\pm2\,{\rm mK}$, and (p-r) $108\pm2\,{\rm mK}$.
In the first two columns, the symbols with errorbars represent the experimental data, and the thick, colored lines represent the corresponding simulation results. The shaded areas represent the error of the numerical estimates due to the uncertainties in the fit parameters.
In order to visualize the total force field $F$ resulting  from the electrostatic repulsion and the CCFs (i.e., from the potential $V_{\rm es} + V_{\rm C}$), in the third column we plot within  the $\Delta x$-$\Delta y$ plane ($\Delta x = x_2 - x_1$ and $\Delta y = y_2-y_1$) the corresponding vector, indicating repulsion by blue arrows and attraction by red ones. Their length corresponds to $4.6\times 10^{-3}\log_{10} (|F|/10 {\rm fN})$ for the scale used.
}
\label{fig6}
\end{figure*}

In the first two columns of Fig.~\ref{fig6}, we report the values of the experimental and simulated drift velocities $v_{\parallel}$ and $v_{\perp}$ as functions of $r/d$ and for decreasing  values of $\Delta T$ from top to bottom. 
The theoretical values of the drift velocities are calculated by employing Eqs.~\eqref{eq:def-vpp} and \eqref{eq:def-vpe} for the simulated trajectories. These trajectories are obtained from Brownian dynamics simulations of two particles interacting via the total potential $V = V_{\rm C} + V_{\rm es}$. We simulate a Langevin equation along the lines of Ref.~\cite{volpe2013simulation}, with the diffusion coefficients $D_{\perp,\parallel}$ 
following from Eq.~(5.5) and Fig.~3 in Ref.~\cite{batchelor1976brownian} (see also Fig.~\ref{fig5} here) and based on the value of $D_0$ as given by  Eq.~\eqref{eq:D0}.
The parameters $\xi$, $\ell_{\rm D}$, and $\rho_{\rm es}$ are fixed to their best-fit values obtained by fitting the initial distribution $P_{eq}(r)$ as described above [see Eqs. \eqref{eq:VC} and \eqref{eq:Ves}].

At large values of $\Delta T$ (Figs.~\ref{fig6}a and \ref{fig6}d), the parallel drift velocity  $v_\parallel$ is positive at small values of $r/d$ because, on average, the particles are pushed away from each other by the dominating repulsive electrostatic potential $V_{es}$. Increasing the value of $r/d$, $v_\parallel$ rapidly vanishes because, correspondingly, the electrostatic repulsion decays exponentially on a scale set by $\ell_{\rm D} \simeq 13\,{\rm nm}$, corresponding to $\ell_{\rm D}/d \simeq 6\times 10^{-3}$ on the scale of the plot.
Upon decreasing $\Delta T$, as in Figs.~\ref{fig6}g and \ref{fig6}j, $v_\parallel$ becomes negative within a certain range of values of $r/d$.  At these distances, the particles move on average towards each other due to the attractive critical Casimir interaction $V_{\rm C}$, which competes and eventually overcomes the electrostatic interaction $V_{\rm es}$. 
However, at smaller values of $r/d$, $V_{\rm es}$ dominates and $v_\parallel$ is no longer negative. 
If $\Delta T$ is reduced further, $v_\parallel$ becomes quickly more negative (Figs.~\ref{fig6}m and \ref{fig6}p), because the attractive critical Casimir interaction is so strong that it moves the particles towards each other until their velocity vanishes at contact.   
At larger distances, instead, $v_\parallel$ vanishes and the particles undergo Brownian diffusion.
We note that this range of distances can actually be explored only via numerical simulations with sufficiently high statistics. In the experiment, instead, the particles turn out to stick almost always together, and they explore the very limited range of distances indicated on the solid horizontal axes.

The experimental and numerical determination of the orthogonal component $v_\perp$ of the drift velocity (Eq.~\eqref{eq:def-vpe}) are reported in the second column of Fig.~\ref{fig6}. Here, $v_\perp$ vanishes in all the cases investigated because all the forces at play in the present experiment act along the direction which connects the centers of the particles. Accordingly, $v_\perp$ shows no temperature dependence.

The third column of Fig.~\ref{fig6} reports the total force field (resulting from the sum of the electrostatic force and of the attractive CCF) 
in the $xy$-plane used in the numerical simulation. 
The length of the arrows corresponding to each point in that plane is proportional, for the purpose of visualization, to the logarithm of the magnitude of the total force. Blue arrows indicate repulsive forces whereas red arrows indicate attractive ones.

It is noteworthy that the agreement observed in Figs.~\ref{fig5} and \ref{fig6} between the experimental and simulated data, confirms the reliability of the model we have used.

\begin{figure*}[t]
\includegraphics[width=0.95\textwidth]{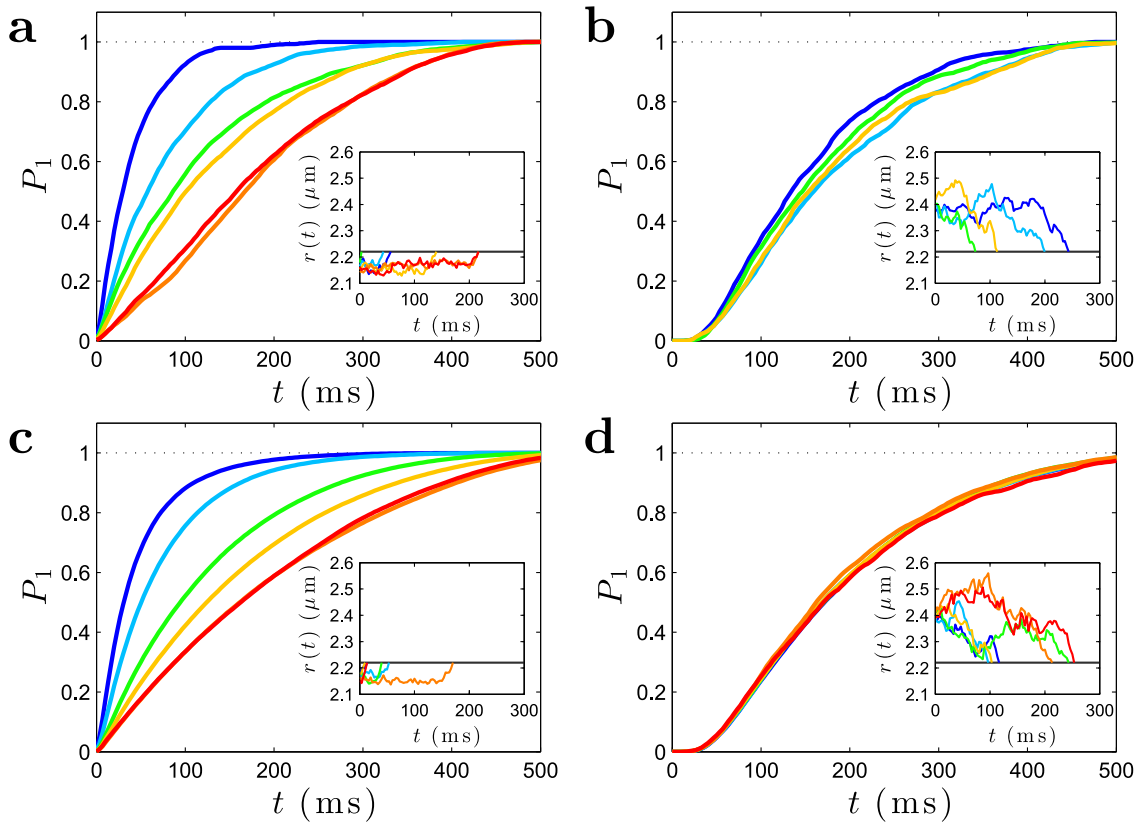}
\caption{Cumulative probability $P_1(t)$ of first-passage times at a certain inter-particle distance as a function of time for various temperatures. 
(a-b) Experimentally and (c-d) numerically determined probability distribution $P_1(t)$ to reach the reference distance $r_{\rm ref}=2.22\,{\rm \mu m}$ for the first time earlier than a given time $t$, when the particles starts from an initial distance (a and c) $r_{\rm in} = 2.16\,{\rm \mu m} < r_{\rm ref}$ and (b and d) $r_{\rm in} = 2.40\,{\rm \mu m} >  r_{\rm ref}$. 
The lines of various colors refer to $\Delta T=456\pm2\,{\rm mK}$ (blue),  $0.2730\pm2\,{\rm mK}$ (light blue), $200\pm2\,{\rm mK}$ (green), $163\pm2\,{\rm mK}$ (yellow), $127\pm2\,{\rm mK}$ (orange), and $108\pm2\,{\rm mK}$ (red).
The insets in the various panels show, on the same scale, representative trajectories of $r(t)$ for various temperatures (the horizontal solid lines correspond to $r_{\rm ref}$). 
In panel (d) and on that scale, the various curves are almost indistinguishable.}
\label{fig7}
\end{figure*}

\emph{First passage times --}  In order to highlight the effect of CCFs on the dynamics of the two colloidal particles, we consider the first-passage time $t_1$, i.e., the time it takes that the particles reach for the first time a reference separation $r_{\rm ref}$, starting from a certain initial distance $r_{\rm in}$. 
Heuristically, this quantity provides a measure of how much the interaction effectively speeds up or slows down the relative diffusion process of the two particles, $t_1$ being the \emph{minimal time} required by the colloids to realize a certain configuration.

The first-passage time is a random variable which changes for each realization of the diffusion process. Accordingly, it can be characterized by its cumulative probability distribution $P_1(t)$ that $t_1$ is smaller than a given time $t$, which depends on the choice of $r_{\rm in}$ and $r_{\rm ref}$. 
(In order to smooth out statistical fluctuations of the experimental data, it is convenient here to focus on the cumulative distribution of $t_1$ instead of its probability density $p(t_1)$, which can in principle be obtained as $p(t_1)= ({\rm d}P_1(t)/{\rm d}t) |_{t=t_1}$.)
In order to determine $P_1$, for each value of $\Delta T$ and each repetition of the blinking process, in which the particles are initially separated by a distance $r_{\rm in}$, we measure the time $t_1$ it takes them to reach the separation $r_{\rm ref}$ for the first time, while the optical traps are turned off. Based on these data, $P_1(t)$ at a certain time $t$ is determined by the ratio between the number of occurrences for which $t_1 < t$ and the total number of collected data.
The experimental results are presented in Figs.~\ref{fig7}a and \ref{fig7}b for $r_{\rm in} = 2.16\,{\mu m}< r_{\rm ref}$ and $r_{\rm in}=2.40\,{\rm \mu m}>r_{\rm ref}$, respectively, with $r_{\rm ref}=2.22\,{\rm \mu m}$ in all cases.


First we consider the case of the first passage time from $r_{\rm in} =  2.16\,{\mu m}$ to $r_{\rm ref}=2.22\,{\rm \mu m}$, as shown in Fig.~\ref{fig7}a.
Far from criticality ($\Delta T = 456\pm2\,{\rm mK}$), the repulsive electrostatic interaction dominates for $r<r_{\rm ref}$ and therefore, as the particles are pushed away from each other,  $P_1(t)$ rapidly reaches  its maximal value 1. However, upon approaching criticality, the increasingly strong, attractive critical Casimir interaction, acting for $r < r_{\rm ref}$, effectively slows down the separation of the particles, so that $P_1(t)$ approaches its maximum value 1 only at long times. 

Setting the initial condition outside the range of action of the CCFs, i.e., for $r_{\rm in} = 2.40\,{\rm \mu m}$,  $P_1(t)$ has a significantly less pronounced dependence on $\Delta T$, at least within the range of parameters explored here (Fig.~\ref{fig7}b).
This is due to the fact that the temperature dependent CCFs are actually negligible for $r>r_{\rm ref}$. For this range of distances, a  behavior similar to that reported in Fig.~\ref{fig7}a can be obtained only for values of the correlation length $\xi$ significantly larger than those achieved here. 

The results of corresponding Langevin-dynamics simulations are presented in Figs.~\ref{fig7}c and \ref{fig7}d. These simulations are in very good agreement with the experimental data presented in Figs.~\ref{fig7}a and \ref{fig7}b, respectively. 
This agreement further validates our simulation model, which is based on an interaction potential $V(\rho)=V_{\rm C}(\rho)+V_{\rm es}(\rho)$, as function of the surface-to-surface distance $\rho$ between the particles, with a diffusion term described according to Eqs.~\eqref{eq:def-Dpp} and \eqref{eq:def-Dpe} \citep{paladugu2016nonadditivity}.

\begin{figure*}[t]
\includegraphics[width=0.95\textwidth]{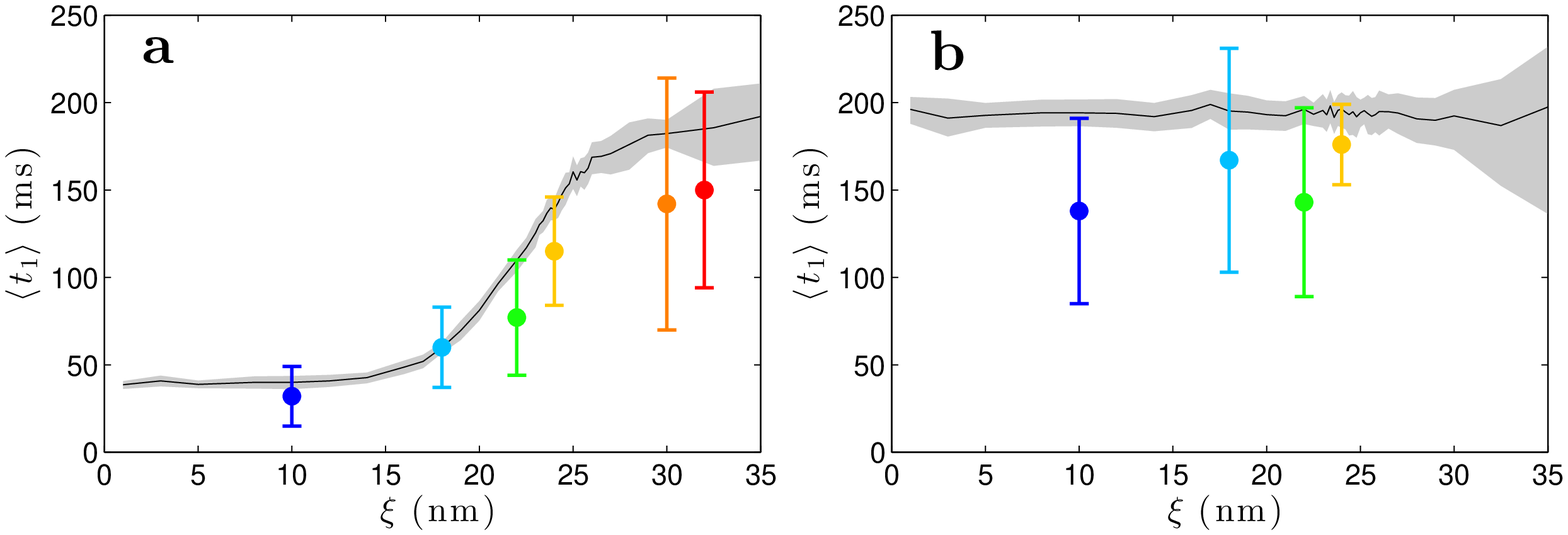}
\caption{Mean first-passage time $\langle t_1\rangle$, as a function of the correlation length $\xi$, to reach for the first time the reference distance $r_{\rm ref}=2.22\,{\rm \mu m}$ from the initial distance (a) $r_{\rm in}=2.16\,{\rm \mu m}$ and (b) $2.40\,{\rm \mu m}$.
Symbols with errorbars represent the experimental data while black solid lines with greyly shaded area correspond to simulation data and their uncertainties. 
The color code for the temperatures is the same as the one used in Figs.~\ref{fig5} and \ref{fig6}. Here $\langle t_1\rangle$ is reported as function of the associate correlation lengths $\xi$ instead of $\Delta T$; specifically,  $\xi = 10\,{\rm nm}$ (blue), $18\,{\rm nm}$ (light blue), $22\,{\rm nm}$ (green), $24\,{\rm nm}$ (yellow), $30\,{\rm nm}$ (orange), and $32\,{\rm nm}$ (red).}
\label{fig8}
\end{figure*}

The results reported in Fig.~\ref{fig7} can be made more quantitative by calculating the mean first-passage time $\langle t_1 \rangle = \int_0^{+\infty}{\rm d} t_1 \, t_1 p(t_1) = \int_0^{+\infty}{\rm d} t \, [1-P_1(t)]$, 
which is reported in Fig.~\ref{fig8} as a function of the correlation length $\xi$ corresponding to the values of $\Delta T$ used in Fig.~\ref{fig7}.
For $r_{\rm in}=2.16\,{\rm \mu m}$ and $r_{\rm ref}=2.22\,{\rm \mu m}$ (Figs.~\ref{fig7}a and \ref{fig7}c), $\langle t_1 \rangle$ increases upon increasing the correlation length $\xi$ (Fig.~\ref{fig8}a), because the attractive interaction due to $V_{\rm C}$ slows down the diffusion of the particles.
Instead, for $r_{\rm in}=2.40\,{\rm \mu m}$ and $r_{\rm ref}=2.22\,{\rm \mu m}$ (Figs.~\ref{fig7}b and \ref{fig7}d), no significant dependence of $\langle t_1 \rangle$ on $\xi$ is observed (Fig.~\ref{fig8}b), because the CCFs are negligible for $r\gtrsim r_{\rm ref}$ within the range of values of $\xi$ explored here. 

\emph{Conclusions --}  We have shown that by using blinking optical tweezers it is possible to investigate the dynamics of a pair of colloidal particles, dispersed in a critical, binary liquid solution of water and 2,6-lutidine, in the absence of optical potentials and, thus, occurring under the influence of their effective inter-particle interaction only.
Digital video microscopy facilitated to track the positions of the particles and to determine the effects of the CCFs on the time evolution of their center-to-center distance $r$ upon approaching the critical temperature of the solvent. 
In order to infer the correlation length $\xi$ of the critical fluctuations and the strength of the corresponding CCF, we have compared the experimental data for the equilibrium distribution of the inter-particle distance in the presence of optical traps with the results of a Monte Carlo integration of the expected Boltzmann distribution. The resulting fitted parameters have been used to perform a simulation of the dynamics of the two interacting colloids immersed in the same solvent.
The very good agreement between the experimental data and the corresponding numerical simulations based on Langevin dynamics has validated the theoretical description of the forces involved and of the dynamics, which does not require accounting for the possible effects of 
retardation \cite{furukawa2013nonequilibrium}.
This holds at least within the range of parameters explored here. The agreement between theory and experiment has also provided informations about the correlation length $\xi$ of the critical fluctuations and about the CCF field, for various temperatures approaching the critical point. 
Moreover, the knowledge of the first passage time relative to a start and a final configuration, in which the particles are essentially fully separated, is crucial for understanding the dynamics and to eventually control the self-assembly process of many colloidal particles. In particular, the model used here can be exploited to create a base protocol for the application and for the fine tuning of CCFs towards their use for nanotechnology. This offers new possibilities for the design and realization of self-assembled nano-structures and for driving nano-devices. 

\begin{acknowledgments}
This work was partially supported by the ERC Starting Grant ComplexSwimmers (grant 
nr.~677511) and by Vetenskapsr{\aa}det (grant nr.~2016-03523). A.~C.~acknowledges partial financial support from TUBITAK (grant nr.~116F111).
\end{acknowledgments}


\begin{thebibliography}{41}
\bibitem{fn1}
In order to illustrate the emergence of these corrections via using a simple example, we assume that the dynamics of the distance $r$ under the sole effect of the interparticle potential $V_{\rm es}+V_{\rm C}$ is described  by the overdamped Langevin equation $\dot r = f(r) + \sqrt{2 {\cal D}(r)}\, \eta(t)$, where $f(r)$ is proportional to the deterministic forces acting on the particles (due to that potential), $\eta(t)$ is a Gaussian noise with $\langle \eta(t)\rangle =0$, and $\langle \eta(t)\eta(t')\rangle = \delta (t-t')$, while ${\cal D}(r)>0$ is the position-dependent diffusion constant. 
%
This stochastic differential equation is interpreted according to the \^Ito convention, while here we neglect the fact that $r$ is actually the projection of the interparticle distance onto the $xy$-plane.  
%
A relatively simple but lengthy calculation shows that by defining $D(r)$ as in Eq.~\eqref{dif}, i.e., from the conditional expectation of the increments $\Delta r$ of the variable $r$ within a time interval of finite duration $\Delta t$, one finds $D(r) = {\cal D}(r) + (\Delta t /2)c_1(r) + {\cal O}((\Delta t)^2)$, where $c_1(r) = f^2(r) + 2 {\cal D}(r)f'(r)+{\cal D}'(r)f(r) + {\cal D}(r){\cal D}''(r)$ is the leading finite-time bias of the estimator $D$ of the actual diffusion ${\cal D}$. 
%
We emphasize that $c_1(r)$ vanishes at large separations because, correspondingly, $f(r)$ generically decays to zero while ${\cal D}(r)$ approaches its bulk, $r$-independent value. In addition, the sign of $c_1(r)$ may change as a function of $r$. In particular, one has $c_1(r)<0$ when $r$ approaches a point $r_0$ of stable mechanical equilibrium [i.e., with $f(r_0)=0$ and $f'(r_0)<0$] with ${\cal D}''(r_0)<0$, as in the case of the theoretical diffusion coefficients reported in Fig.~\ref{fig5}.
\end{thebibliography}
%

\end{document}